\documentclass[preprint,a4paper,showkeys]{revtex4-1}

\usepackage{amsmath}
\usepackage{amssymb}
\usepackage{ulem}

\usepackage{epsfig,latexsym}

\usepackage{color}

\begin{document}

% %%%%%%%%%%%%%%%%%%%%%%%%%%%%%%%%%%%%%%%%%%%%%%%%%
%
% FRONT MATTER
%
% %%%%%%%%%%%%%%%%%%%%%%%%%%%%%%%%%%%%%%%%%%%%%%%%%

\title{The Electronic Behavior of Zinc-Finger Protein Binding Sites in the Context of the DNA Extended Ladder Model}

\author{Nestor Norio Oiwa}
\email{nestoroiwa@vm.uff.br}
\affiliation{Department of Basic Science, Universidade Federal Fluminense, Rua Doutor S\'{i}lvio Henrique Braune 22, Centro, 28625-650 Nova Friburgo,  Brazil}
\affiliation{Institute for Theoretical Physics, Heidelberg University, Philosophenweg 19, D-69120 Heidelberg, Germany}

\author{Claudette El\'\i sea Cordeiro}
\email{clau@if.uff.br}
\affiliation{Department of Physics, Universidade Federal Fluminense, Avenida Gal. Milton Tavares de Souza s/n, Gragoat\'a, 24210-346 Niter\'oi, Brazil}

\author{Dieter W. Heermann}
\email{heermann@tphys.uni-heidelberg.de}
\affiliation{Institute for Theoretical Physics, Heidelberg University, Philosophenweg 19, D-69120 Heidelberg, Germany}

\date{\today}

%%%%%%%%%%%%%%%%%%%%%%%%%%%%%%%%%%%%%%%%%%%%%%%%%%
%
% ABSTRACT
%
%%%%%%%%%%%%%%%%%%%%%%%%%%%%%%%%%%%%%%%%%%%%%%%%%%
\begin{abstract}
The eukaryotic Cys2His2 zinc finger proteins bind to DNA ubiquitously at
highly conserved domains, responsible for gene regulation and the spatial
organization of DNA. To study and understand the zinc finger DNA-protein
interaction, we use the extended ladder in the DNA model proposed by
Zhu, Rasmussen, Balatsky \& Bishop (2007) \cite{Zhu-2007}. Considering
one single spinless electron in each nucleotide $\pi$-orbital along a
double DNA chain (dDNA), we find a typical pattern for the
{\color{black} bottom of the occupied molecular orbital (BOMO)}, 
highest occupied molecular orbital (HOMO) and lowest unoccupied
orbital (LUMO) along the binding sites. We specifically looked at two 
members of zinc finger protein family: specificity protein 1 (SP1) and
early grown response 1 transcription factors (EGR1). When the valence
band is filled, we find electrons in the purines along the nucleotide
sequence, compatible with the electric charges of the binding amino
acids in SP1 and EGR1 zinc finger.
\end{abstract}


\maketitle

%%%%%%%%%%%%%%%%%%%%%%%%%%%%%%%%%%%%%%%%%%%%%%%%%%
%
% INTRODUCTION
%
%%%%%%%%%%%%%%%%%%%%%%%%%%%%%%%%%%%%%%%%%%%%%%%%%%
\section*{Introduction}
 
One of the major eukaryotic DNA-protein binding motifs are those related
to zinc fingers (ZF), the key protein family for the chromatin
condensation  as well as the  gene regulation. There are around one
thousand ZF encoding genes \cite{Razin-2012} and ten thousands highly
conserved putative ZF binding sites along the human genome
\cite{Kim-2007,Chen-2012}. The majority of ZF proteins assist
transcription factors, acting as repressors, activators and regulators
\cite{Klug-2010,Razin-2012}. They are responsible for the genome spacial
structure in the DNA loops too, exposing or hiding the genes, and work
as an insulator, avoiding the spread of heterochromatin 
\cite{Ong-2014}. These ZF proteins could mediate long-range chromosomal
interactions in eukaryotic cells, $>$ 100 thousand base pair (bp)
\cite{Yusufzai-2004,Kurukuti-2006,Ling-2006, Feinauer-2013}. However,
the exact relation between the long-ranged correlation in genomic scale
nucleotide sequences ($>$20 thousand bp) and the chromosomal
three-dimensional organization is still not clear
\cite{Peng-1992,Peng-1994,Oiwa-2000,Oiwa-2005, Heermann-2011,
Feinauer-2013} and subject to intense research. Furthermore, since
transcription factors spot specific sequences without the opening of the
double helix, we expect some biological mechanism for probing nucleotide
based on local properties \cite{Oiwa-2007}. To understand this there are
two basic approaches: a polymeric description and by electric charges.
The most common polymeric description considers DNA as a single
one-dimensional strand, explaining the DNA denaturation
semi-analytically  \cite{Peyrad-1989, Macia-2007, Weber-2013}. The
literature also report the mechanical properties of chromosomal fibers
and the long-range nucleotide interaction due to DNA loops, histones and
zinc finger proteins, using Monte Carlo simulation \cite{Langowski-2007,
Bohn-2007, Feinauer-2013, Lei-2015}.  Since electrons play a crucial
role in the DNA-protein interaction, we must consider the DNA from the
electric charge distribution too. 
{\color{black} The electronic nature of DNA is still in debate, but the literature point us some cues about their organization.
The double helices behave as isolants or conductor 
under silver deposition \cite{Braun-1998}, material contaminants} \cite{Fink-1999,Pablo-2000} 
{\color{black} and others environmental conditions  \cite{Taniguchi-2006}.
However, when the conductivity is measures in atmosphere, low vacuum or Tris-HCl buffers, DNA has semiconductor features
with the typical gap between the valence and conductor band in the electronic density of states (DOS)}
\cite{Taniguchi-2006, Tran-2000, Cai-2000, Porath-2000, Yoo-2001, Lee-2002}.
In order to describe this behavior,
ionization models (also known as ballistic, polaron or wire-like charge
transport) have been proposed \cite{Yamada-2004, Oliveira-2006,
Sarmento-2009, Venkatramani-2011, Sarmento-2012}.  The parameters in
ionization models are easily measured, since one just needs to evaluate
the loss of energy when we takes one electron in a neutral molecule. The
lost electron is usually in the highest occupied molecular orbital
valence band (HOMO) and it may easily jump to the lowest unoccupied
molecular orbital in the conductor band (LUMO). But, the literature 
also suggests electronic affinity models, where the energy is described
by the gain of electrons in neutral molecule
\cite{Richardson-2004,Lu-2005, Chen-2007,Chen-2009,Gu-2012}. These
theoretical results usually combine density functional theory and
molecular dynamic simulation. 

In 2007 Zhu et al. joined both molecular ionization and affinity
approaches \cite{Zhu-2007}. This adaptation of the Peyra-Bishop DNA melting
model \cite{Peyrad-1989} describes the nucleotide sequence from their
semi-conductor features, avoiding the heavy computational cost of {\it
ab initio} molecular dynamical simulations. Their work allowed to spot
electronic local density of states (LDOS) in one viral P5 promoter
sequence, connecting LDOS with one specific biological function
\cite{Zhu-2007,Oiwa-2007}. Unfortunately, they did not observe in their
model the gap between HOMO and LUMO in $(C)_n$ as one expects from the
experimental data \cite{Yoo-2001,Shapir-2008}. In our work we fix the
problem of HOMO-LUMO gap, introducing the extended ladder in the model
as suggested by
\cite{Senthilkumar-2005,Mehrez-2005,Zilly-2010,Sarmento-2012}. We also
analyze systematically the DNA-protein binding sites for two
transcription factor proteins:  the human specificity protein
transcription factor 1 (SP1) \cite{Kaczynski-2003} and early grown
response factor (EGR1, aka  Zif268) \cite{Thiel-2002} both localized in
the promoter of a great variety of genes and characterized by a
molecular structure called {\color{black} Cys$_2$His$_2$ zinc finger (ZF). 
The descriptions of ZFs as well as SP1 and EGR1 are in the appendix.}

This paper is organized as follows. First, we describe the extended
ladder model. Then, we test the model, studying the electronic behavior
of $(C)_n$ and $(T)_n$ sequences. We replace one nucleotide in order to
understand the electronic interactions along the DNA chain. After this,
we analyze systematically the SP1 and EGR1 binding sequences and report
strand dependence and independence results.

%%%%%%%%%%%%%%%%%%%%%%%%%%%%%%%%%%%%%%%%%%%%%%%%%%
%
\section*{The Model}
%
%%%%%%%%%%%%%%%%%%%%%%%%%%%%%%%%%%%%%%%%%%%%%%%%%%

In this paper, we consider one double DNA chain with $n$ base pairs,
totaling $2n$ nucleotides, Fig. \ref{modelo01}(d). In
reality our model does not consider nucleotides, but {\bf nucleosides},
i.e. the {\bf nucleotide} with the phosphate group. However, we will
call nucleosides nucleotides in this work in order to simplify the
nomenclature. The electronic behavior of the spinless free electron of
the $\pi$-orbital of the nucleotide is described using the same
Hamiltonian as in
\cite{Zhu-2007},\begin{equation}
 H=H_e+H_{eb}+H_b. 
      \label{modelo02}
\end{equation}
The first term in Eq. \ref{modelo02} is 
{\color{black} given by,}
\begin{eqnarray}
H_e&=&\sum_{i=1}^{2n}\epsilon_iC_i^{\dagger}C_i
     + ( \sum_{i=1}^{n-1}t_{2i-1,2i+1}C_{2i-1}^{\dagger}C_{2i+1} \
     +\sum_{i=1}^{n-1}t_{2i,2i+2}C_{2i}^{\dagger}C_{2i+2} \nonumber\\
     &&+\sum_{i=1}^{n-1}t_{2i-1,2i}C_{2i-1}^{\dagger}C_{2i}
     +\sum_{i=1}^{n-1}t_{2i-2,2i+1}C_{2i-2}^{\dagger}C_{2i+1} )+H.c.\\ \nonumber\label{He01}
\end{eqnarray}
where $C^{\dagger}_i$ and $C_i$ are the electron creation and
annihilation operators at site $i$, $\epsilon_i$ is the on-site ionization
energy, $n$ is the number of nucleotides and $t_{ij}$ is the electron
hopping rate between nucleotides $i$ and $j$,  Here, we are using the
extended ladder, where we duplicate the one dimensional lattice in
\cite{Zhu-2007} and include the interstrand hopping
\cite{Senthilkumar-2005,Zilly-2010,Sarmento-2012}. The structure of the
ladder considers the long-distance charge and hole transport along dDNA
\cite{Jortner-1998,Giese-2000,Bixon-2002,Senthilkumar-2005} The second
term in Eq. \ref{modelo02} represents the coupling between the free
electron and the nucleotide displacement field,
\begin{equation}
    H_{eb}=\alpha_v\sum_{i=1}^{2n}y_iC_i^{\dagger}C_i\label{Heb01}
\end{equation}
where $y_i$ is the displacement (dark dotted line) of the electronic
cloud from the equilibrium in the nucleotide (light dotted line), Fig.
\ref{modelo01}(d). The last term $H_b$ represents the interaction of the
electron with the nucleotide:
\begin{equation}
   H_b=\sum^{2n}_{i=1}[D_i ( e^{-a_iy_i}-1)^2+\frac{k_v}{2}(1+\rho e^{-\alpha(y_i+y_{i+1})})(y_i-y_{i-1})^2],\label{Morse01}\\
\end{equation}
where $D_i$ and $a_i$ are parameters of the Morse potential, $k_v$ is
the spring constant of the anharmonic interaction between two contiguous
base-pairs. $\rho$ and $\alpha$ are the parameters for modifying $k_v$
in order to evaluate long-range cooperative electronic behavior
\cite{Zhu-2007}.

We study the electronic part $H_e$ and $H_{eb}$ of the Hamiltonian in
Eq. \ref{modelo02} computing the eigenvalue $E_k$ and eigenvectors
$\phi_i^k$, $i,k=1,...,2n$, of the matrix 
\begin{equation} 
H_{e+eb} = \left| \begin{array}{cccccccc}
\epsilon_1+\alpha_v y_1 & t_{1,2} & t_{1,3} & t_{1,4} & 0 & 0 & 0 & \ldots \\
t_{2,1} & \epsilon_2+\alpha_v y_2 & t_{2,3} & t_{3,4}& 0 & 0  & 0 & \ldots  \\
t_{3,1} & t_{3,2} & \epsilon_3+\alpha_v y_3 & t_{3,4} & t_{3,5} & t_{3,6} & 0 & \ldots  \\
t_{4,1} & t_{4,2} & t_{4,3} & \epsilon_4+\alpha_v y_4 & t_{4,5} & t_{4,6} & 0 & \ldots \\
0 & 0 & t_{5,3} & t_{5,4}  & \epsilon_5+\alpha_v y_5 & t_{5,6} & \ldots & \ldots \\
0 & 0 & t_{6,3} & t_{6,4} & t_{6,5} & \epsilon_6+\alpha_v y_6  & \ldots & \ldots \\
0 & 0 & 0 & 0 & \vdots &\vdots &\ddots &\ddots \\
\vdots &\vdots & \vdots &\vdots & \vdots &\vdots &\ddots &\ddots \\
\end{array} \right|.\label{matHeeb}
\end{equation}
This matrix is similar to the one suggested in ~\cite{Zilly-2010,Sarmento-2012}, except for
the electron base component $H_{eb}$.

In order to estimate $y_i$, we consider the self-consistency condition, given by
\begin{equation}
<\frac{\partial H_b}{\partial y_i}+\frac{\partial H_{eb}}{\partial y_i}>=0,\label{selfconsist01}
\end{equation}
where $<...>$ represent the average over the free electrons in the
system. The iteration method for solving Eqs. \ref{matHeeb} and
\ref{selfconsist01} is described in \cite{Zhu-2007}, and it consists of
the follow procedure. Given a initial condition for $\{y_i\}$, we
diagonalize the matrix \ref{matHeeb} in order to compute the electronic
occupation in each site $<n_i>$, where $
n_i=\sum_{k=1}^{n_e}|\phi_i^{k}|^2$ and $n_e$ is the number of electrons
in the system. This set of $<n_i>$ will be used in the Langevin equation
calculated from Eq. \ref{selfconsist01}. We update the values of
$\{y_i\}$, using fourth-order Runger-Kutta method for the Langevin
equation. The new $\{y_i\}$ set is inserted again in the matrix of Eq.
\ref{matHeeb}. We repeat the iteration until we achieve the minimum
local adiabatic electronic and structural configuration. The computation
were done using R with the package deSolve for the Runger-Kutta
algorithm \cite{Soetaert-2010}. 
The choices of the model parameters are in the appendix.

In this work, we estimate {\color{black} spatial distribution of electrons}, energy level and displacement field only
considering $n_e=n$. Thus, the valence band is always filled
with electrons and the conductor band is empty. Our model does not have
periodic boundary condition. So, the elected regions for analysis must
be larger in order to avoid boundary effects. We analyze only nucleotide
sequences at least 10bp distant from the beginning and ending of the
sample.

We apply the proposed model in  poly(C)-poly(G) and
poly(T)-poly(A) sequences with 63 base pairs in order to understand the
behavior of the electrons dispersed along the DNA chain.
  
\begin{figure}[th]
     \epsfxsize =\textwidth
     \begin{center}
          \leavevmode
          \epsfbox{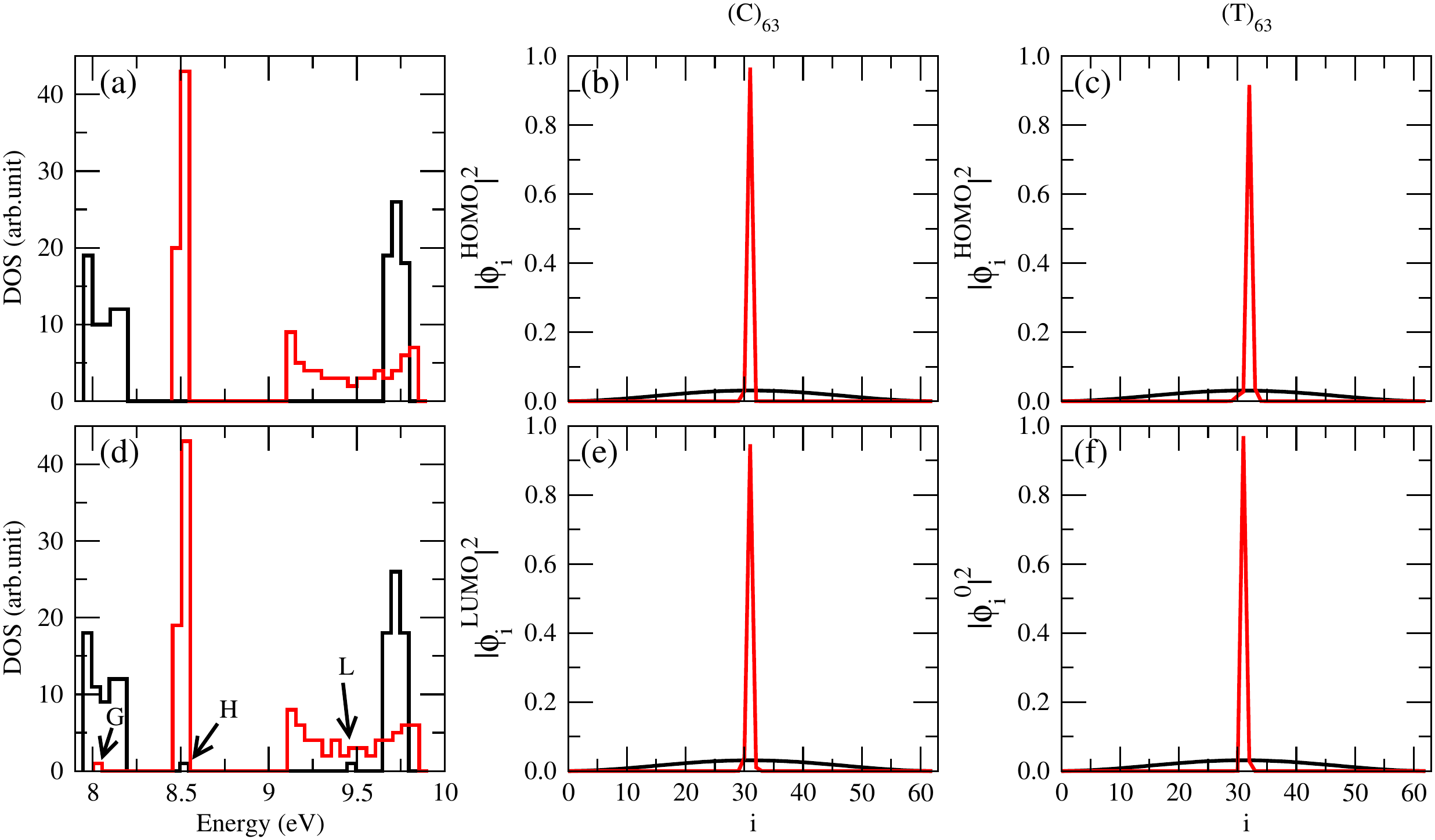}
     \end{center}
     \caption{
     (a) The electronic density of states (DOS) for $(C)_{63}$ in black
     lines and $(T)_{63}$ in red lines. (d) the same as in (a), except
     that the sequence has one $C$ or $T$ in $i=32$.  In (d) {\color{black} BOMO}
      for $(T)_{63}$ with one replaced C in the position 32
     is pointed as G, and the HOMO and LUMO energetic level for
     $(C)_{63}$ with $T$ in $i=32$ are respectively indicated by H and
     L. The electronic cloud for HOMO $|\phi_i^{\mbox{\small HOMO}}|^2$
     (b) and LUMO $|\phi_i^{\mbox{\small LUMO}}|^2$(e) for $(C)_{63}$
     (black lines) and the same sequence with $T$ in $i=32$ (red lines).
     The electronic cloud for HOMO $|\phi_i^{\mbox{\small HOMO}}|^2$ (b)
     and {\color{black} BOMO} $|\phi_i^{\mbox{\small 0}}|^2$ (f) for $(T)_{63}$
     (black lines) and the same sequence with $C$ in position 32 (red
     lines). }
     \label{C63T63spectra}
\end{figure}

{\color{black} According to the literature \cite{Mehrez-2005,Zilly-2010,
Sarmento-2012,Wang-2004}, we} do expect a gap in the energy band in
the test sequences $(C)_{63}$ and $(T)_{63}$ as can be seen in Fig.
\ref{C63T63spectra}(a). 
The gap between the valence and conductor band
in $(T)_{63}$ is narrower than in $(C)_{63}$. 
Furthermore, the {\color {red} gap} 
of the pure $(C)_{63}$
sequence can be {\color{black} modulated,} 
when we introduce one single $T$ in the
position 32. One HOMO and LUMO appear in the gap of the energy band,
marked as H and L in Fig. \ref{C63T63spectra}(d). Moreover, the HOMO and
LUMO electronic cloud, dispersed in pure $(C)_{63}$, black lines in Fig.
\ref{C63T63spectra}(b,e), becomes localized in the introduced $T$ (red
lines in Fig. \ref{C63T63spectra}(b,e)). We remark that the electronic
cloud of HOMO is dispersed in a pure $(C)_{63}$. Thus, thymines and
adenines are related with LUMOs and cytosine and guanines are linked
with the localization of
{\color{black}the bottom of occupied molecular orbital (BOMO).} 

On the other hand, when we substitute one thymine by cytosine in a
$(T)_{63}$, the HOMO electronic cloud will be localized in the replaced
nucleotide, Fig. \ref{C63T63spectra}(b). Moreover, the eigenvalue
related to this electronic state remains in the valence band with values
8.05$\pm$0.01 eV, G in Fig.  \ref{C63T63spectra}(d). Furthermore,  
{\color{black} the electronic distribution of BOMO}
will be positioned over the cytosine too. The CG rich
domains usually 
are related to 
{\color{black} BOMO}. We do not observe any alterations
in the conductor band for a $(T)_{63}$ with and without the replacement.
%\sout{Finally, the sequence $(T)_{63}$ is a n-type semiconductor \cite{Yoo-2001}.}
  
%%%%%%%%%%%%%%%%%%%%%%%%%%%%%%%%%%%%%%%%%%%%%%%%%%
%
% RESULTS
%
%%%%%%%%%%%%%%%%%%%%%%%%%%%%%%%%%%%%%%%%%%%%%%%%%%

\section*{The Electronic Density of State in SP1 and EGR1 transcription factor}

We apply the procedure described in the previous section and we 
estimate the eigenvalues $E_k$ and eigenvectors $\phi_i^k$ {\color{black} of the total Hamiltonian} in Eq.
\ref{modelo02}
for the sequences in Table 2. The criteria for the sequence 
selection as well as the method for nucleotide alignments are  described in the appendix. 
The alignments in Table 2 are in agreement with the consensus sequence in the
literature: 5'-ggggcgggg-3' \cite{Klug-2010,Skerka-1995, Yao-1997,
Wong-2002, Iavarone-1999, Schultz-2003, Zhang-2005} and
5'-gcgggggcg-3' \cite{Wolfe-1999, Skerka-1995, Yao-1997,Petersohn-1995,
Mechtcheriakova-1999} for SP1 and EGR1, respectively.

\begin{figure}[th]
     \epsfxsize =\textwidth
     \begin{center}
          \leavevmode
          \epsfbox{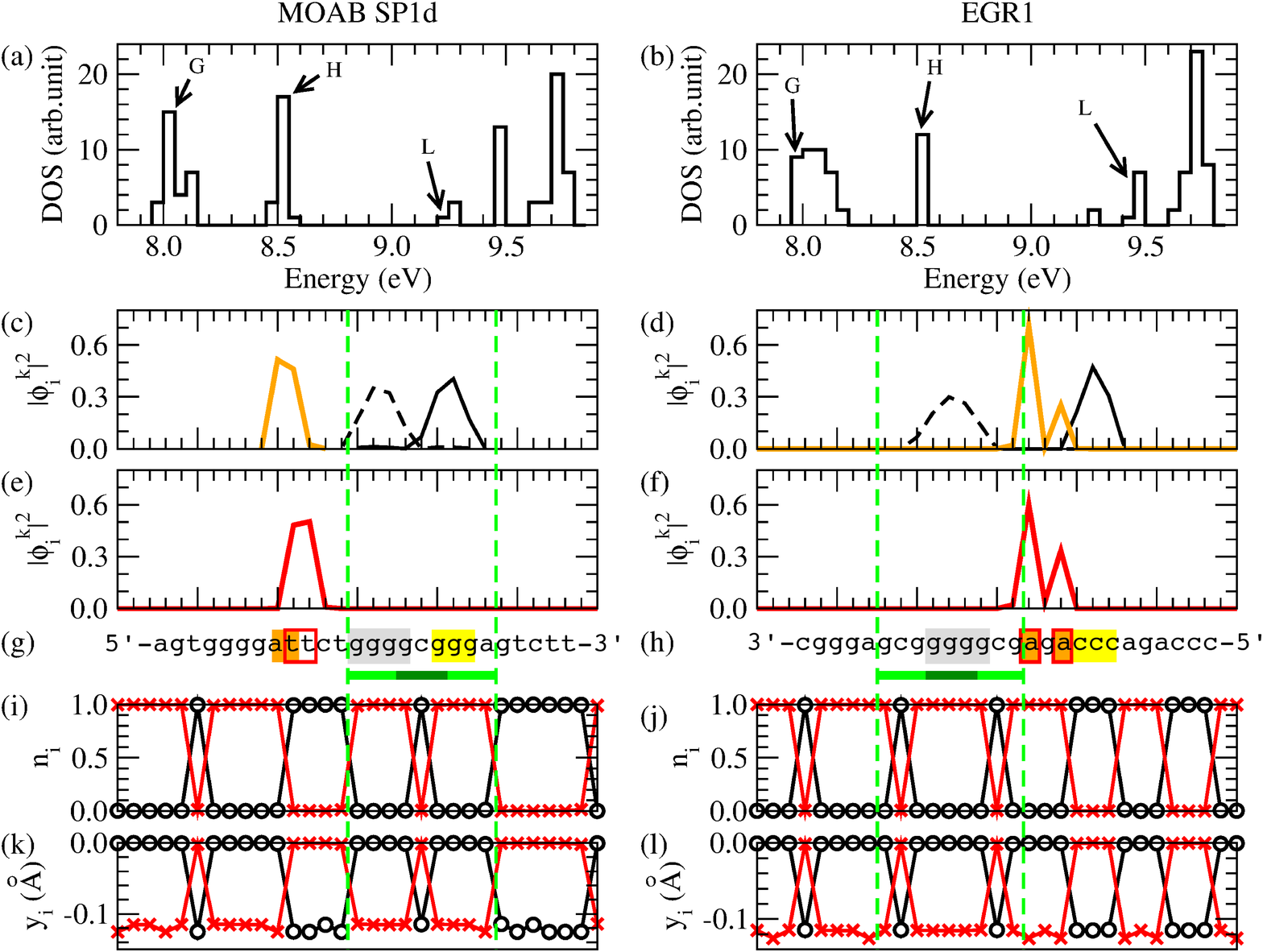}
     \end{center}
     \caption{
     The results of SP1d and EGR1 binding sites for MOAB
     \cite{Wong-2002} and EGR1 genes \cite{Thiel-2002} are respectively
     in the left and right columns. We remark that the EGR1
     transcription factor can bind in the promoter of his own gene
     \cite{Thiel-2002}. (a) and (b) are the density of states (DOS),
     where {\color{black} BOMO}, HOMO and LUMO energy level of the zinc
     finger binding site are indicated by G, H and L. (c) and (d) are
     the probability $|\phi_i^k|^2$ of {\color{black} BOMO} (dashed and solid black lines) and HOMO electrons (orange line). 
     {\color{black} BOMO} $E_0$ in the valence band are degenerated and value
     correspondingly $8.00\pm0.01$eV and $7.98\pm0.02$ eV for SP1d and
     EGR1. $E_{\mbox{\small HOMO}}$ are $8.52\pm0.01$eV in both (c) and
     (d). The electronic clouds $|\phi_i^k|^2$ of LUMO in the conductor
     band are in (e) with $E_{\mbox{\small LUMO}}=9.28\pm0.01$eV and (f)
     with $E_{\mbox{\small LUMO}}=9.45\pm0.01$eV. The nucleotide
     sequences are given in (g) and (h), where we underline the 1-3ZF
     binding consensus sequence in light and dark green lines
     \cite{Klug-2010,Thiel-2002,Skerka-1995,Yao-1997, Wong-2002,
     Iavarone-1999,Schultz-2003,Zhang-2005,Petersohn-1995,
     Mechtcheriakova-1999,Wolfe-1999}. 
     We remark that the MOAB SP1d is in reverse complementary
     direction, the EGR1 reading sequence is in the complementary strand.
     The nucleotides with at least 10\% probability of finding {\color{black} BOMO} electrons are in gray and yellow. The HOMO and LUMO
     nucleotides with $|\phi_i^k|^2\geq 0.1$ are in orange and marked
     with red bordered boxes, respectively. (i) and (j) are the
     probability for the electronic presence in the direct strand (black)
     and the complementary strand (red), when the valence band is
     completely filled, $n_e=n$. (k) and (l) are the field displacements
     $y_i$ in the Morse potential with $n_e=n$ for the direct strand
     (black) and for the complementary strand (red).}
     \label{moabegr}
\end{figure}

Fig.  \ref{moabegr} shows
a typical set of results for the SP1 binding site of the gene MOAB and
EGR1 binding site of the gene EGR1. The nucleotide sequence of MOAB SP1
is in reverse complementary reading direction and EGR1 is in
complementary strand, Fig. \ref{moabegr}(g).

Although we have $2n$ eigenvalues and eigenvectors, each one related
with one of $2n$ nucleotides 
of the system, the relevant electrons for the 
binding sites are those linked with
{\color{black} BOMO}, HOMO and LUMO, respectively noted as G, H and L in
the density of states Fig.\ref{moabegr}(a,b).

We start with the analysis of the position of {\color{black} BOMOs} 
looking for the values of $|\phi_i^k|^2$ with $k$ close to 1.
When we consider $n=50$ as in MOAB and EGR1, the analysis of the first 8 eigenvectors are
usually sufficient to spot the relevant ones. The electronic cloud
$n_i$, $0\leq n_i\leq 1$, Eq. \ref{modelo01}, is strand dependent, but
we do not observe any strand related pattern for individual electrons.
Thus, we sum the  probabilities of the direct and the complementary
strands to find the local electronic cloud. 
{\color{black} BOMOs} could be
degenerated in many electrons along the nucleotide sequence, but we
should focus just in those around the binding sites, yellow and black
lines in the valence band $|\phi_i^k|^2$,
Fig.\ref{moabegr}(c,d). Note that the sum of these two degenerated
{\color{black} BOMOs}  $\sum_k|\phi_i^k|^2\delta(E_0-E_k)$ will result in the
LDOS of the binding site, which is proportional to the differential
tunneling conductance \cite{Zhu-2007}. At low temperature, this quantity
could be measured by scanning tunneling microscope (STM)
\cite{Shapir-2008}. The zinc fingers of SP1 and EGR1 transcription
factors act as tips of an STM, scanning binding sites along the DNA chain.
Finally, we mark the nucleotides with at least 10\% of probability {\color{black} of electronic presence}
 in gray and yellow in Fig.\ref{moabegr}(g,h).

The procedure for localizing the electronic cloud associated with HOMO
and LUMO is very similar to spot 
{\color{black} BOMO probability distributions}, except that $k$ of HOMO
and LUMO are close $n$. 
In order to find the electronic clouds, we
need to consider $k$ from 46 to 50 for HOMO and 51 to 54 for LUMO,
when $n=50$. The electronic cloud associated with LUMO is always close
to the HOMO, with a maximum of $\pm 6$bp distance. Since the probability of
finding one HOMO or LUMO electrons are strand independent, we add both
direct and complementary strand $|\phi_i^k|^2$ in Fig\ref{moabegr}(c-f).
The orange lines in Fig. \ref{moabegr}(c,d) and  red lines in Fig.
\ref{moabegr}(e,f) are HOMO and LUMO, respectively. We can also measure
the LDOS of HOMO and LUMO with STM, using the same approach for 
{\color{black} BOMOs}. The nucleotides with at least 10\% of probability of {\color{black} electronic presence}
is denoted by orange and red boxes in  Fig. \ref{moabegr}(g,h).

Now we return to Table 2, where all {\color{black} BOMOs} are marked in gray and
yellow as well as the HOMO and LUMO electrons are in orange and red
boxes. Looking Table 2, the electronic distribution patterns for the
binding sites for SP1 and EGR1 transcriptor factors are clear.

In the case of SP1, 
{\color{black} BOMO clouds} are over the first
(5'-ggg-3') and third triplets (5'-ggg-3') of the consensus sequence,
light green in Table 2. These triplets are related with the first and
third ZF binding positions of SP1. Moreover, the first 
{\color{black} BOMO} electronic cloud has values
from 4 to 5 bp, while the second ranges from 2 to 4bp. The eigenvalue of
{\color{black} these BOMOs} values $7.98\pm0.05$eV. The energy level
of  HOMO electrons are fixed at $8.52\pm0.02$eV and the electronic cloud
sizes spans between 1 and 2 bp. We observe some fluctuation in the
eigenvalue in LUMO for SP1, which values $9.3\pm0.1$eV. The LUMO
electrons envelop 2 to 5 base pairs. The positions of HOMO and LUMO
associated electrons are always before the first electron and these
electrons are placed from -12 to 1.

For the EGR1 the first 
{\color{black} BOMO} spans from the position 3
to 7 over the second triplet (5'-ggg-3'), and the probability in finding
this particular electron spans over 2 or 4 bp. The second triplet is the
binding site of the second ZF of the early grown response protein 1. The
second electron is after the second triplet and is dispersed between the
nucleotide position 7 to 15, covering the third triplet. The electronic
cloud size ranges from 2 to 4bp. {\color{black} All BOMO} energies in Table 2
values $7.99\pm0.03$eV. The HOMO and LUMO electronic cloud is over the
second electron. All $E_{\mbox{\small HOMO}}$ in Table 2 value
$8.52\pm0.01$eV and the HOMO related electron{\color{black}ic clouds} have a length of 1 or 2
base pair. The LUMO energies fluctuate with an average value of
$9.4\pm0.1$eV. The LUMO electronic cloud sizes vary from 1 to 6bp and
they are in the position from 10 to 20.

Considering the 
HOMO and LUMO {\color{black} distributions}, we believe that they may play
some role in SP1 and EGR1 binding. These proteins bind DNA, embracing
the major grove of the double helix as guide. In the case of SP1, the
head may interact with nucleotides between position -11 to -1. The
behavior for EGR1 is more elusive because HOMO and LUMO are completely
disperse over the nucleotides 5 to 20. Despite the description
emphasizing the similarity between the ZF and nucleotide interaction in
the literature over the consensus nucleotides \cite{Klug-2010}, the
mechanisms of protein attachment in EGR 1 and SP1 are not the same
\cite{Wolfe-1999,Nolte-1998}.

The HOMO and LUMO electron{\color{black}ic clouds} are frequently overlapped, and the main
reason for electrons of HOMO and LUMO  are always in adenine and thymine
rich sequences is as follows. The electrons from the highest occupied
molecular orbital in the valence band should move to the nearby lowest
unoccupied molecular orbital in the conductor band, when the system is
disturbed. And the easiest way for this movement is placing the electron
in regions with higher 
{\color{black} excitability}, i.e. AT rich domains. We may
conjecture that this jump of the electron in the HOMO to the LUMO has a
unknown role in the transcriptor factor SP1 and EGR1.

On the other hand, the less mobile electrons are those in the CG rich
domain, since 
they are at the bottom of the DOS. 
So, we expect to spot {\color{black} BOMOs}
in CG rich-sequences instead of AT rich-regions as we
see for $(T)_{63}$ with one $C$ in the position 32, described in the
previously. Furthermore, these 
{\color{black} BOMOs} are degenerated, i.e. all
electrons present the same energy level. Thus, these cytosine and guanine
rich-regions, typical in promoters, are ideal landmarks for SP1 and EGR1
binding sites.
{\color{black}
The absence o excitation in the lowest states is vital for ZF transcription factors, 
because nucleotides with mobile electrons may change 
the position of the beginning of the gene reading, altering the gene expression. 
We are aware about the those eigenvalues between BOMO 
and HOMO, but we still do not find any obvious pattern 
associated with SP1 and EGR1.}

We never observe overposition between the probabilities of 
{\color{black}BOMOs}
and the overlapped LUMO-HOMO electron{\color{black}ic cloud}s using the criteria of a minimum
10\% of the localization probability of one particular electron at the
samples in Table 2.

%%%%%%%%%%%%%%%%%%%%%%%%%%%%%%%%%%%%%%%%%%%%%%%%%%
%
% The strand dependence of the collective electronic probability and the field displacement
%
%%%%%%%%%%%%%%%%%%%%%%%%%%%%%%%%%%%%%%%%%%%%%%%%%%
\section*{The Collective Electronic Behavior}

The electronic probabilities $|\phi_i^k|^2$ of individual electrons,
discussed in the previous section, are strand independent. However, the
collective electronic probabilities $n_i$ and the field displacement
$y_i$ depend of the strand.

When we have $n_e=n$
electrons, we fill only the valence band and
usually observe in all analyzed sequences, Table 2, 100\% of probability
of {\color{black} electronic presence} 
in purine (adenine or guanine) and the absence
of an electron {\color{black} (hole)} in pyrimidines (thymine or cytosine) in agreement with the
DFT analysis \cite{Gu-2012}. Figs. \ref{moabegr}(i,j) are the
probabilities $n_i$ associated with finding one electron in one nucleotide
for the MAOB SP1d and EGR1 binding site sequences. The {\color{black} electronic presence} 
in each purine gives us a new biological interpretation of
Peng et als. contribution \cite{Peng-1992,Peng-1994}. Using exon and
intron rich segments of the eukaryotic genome, they construct a DNA walk
using purine and pyrimidine as criteria for steps. Then they report
self-affine fractality in the walk, showing long-ranged correlation in
the purine and pyrimidine distribution. When we look to Figs.
\ref{moabegr}(i,j), purines and pyrimidines reflect the electronic
distribution along the DNA chain. This electronic distribution is
related to {\color{black} BOMO,} 
HOMO and LUMO 
{\color{black} distributions}, which work as
ZF binding sites, for example. It is important to stress that they report
a self-affine fractal, but not a self-similar fractal. The self-similarity
is related to the palindromic sequences, connected with DNA-loop
structures as tRNA and rRNA \cite{Oiwa-2002,Oiwa-2004}, while
self-affinity is related to introns \cite{Peng-1992}. Furthermore, the
evidence of polynomial long-ranged nucleotide interaction is also
supported by \cite{Oiwa-2000,Oiwa-2005}. In this work the long
contiguous sequences are represented by a sequence of 0 and 1 for
noncoding and coding nucleotides, where noncoding nucleotides are
intergenic regions and introns and coding nucleotides are genes and
regions for metabolic controls. We made an auto-correlation
analysis over the binary sequence and report correlation between two
coding nucleotides at least 20 thousand bp distance apart. 

The existence of long-ranged correlations has another consequence in the
model. The second term in Eq. \ref{Morse01}, the stacking interaction
between adjacent base pair, mimics the bending of DNA as polymer. But,
we will see that the short-ranged $\rho e^{-\alpha(y_i+y_{i+1})}$ in Eq.
\ref{Morse01} does not contribute to the electronic behavior. This term
has an energetic value of the order of $10^{-4}$ eV, when we consider
typical values for the parameters: $y_i\sim -0.1\AA$, $\rho\sim 10$ and
$\alpha \sim 0.35\AA$. On the other hand, the Morse Potential is of the
order of $10^{-2}$eV. The stacking interaction will be relevant only if we
consider $\rho>100$, but such high experimental values for $\rho$ is
unlike.  This short-ranged element of the model comes from the DNA
melting problem, where the interstrands binding of the double helix may
open \cite{Peyrad-1989}. In this case, the
short-ranged element is important, since it is easier to open the dDNA
when the neighbor bp is already open. Moreover, $y_i$ represent the
displacement field of the electronic cloud of the hydrogen bonds between
nucleotides in the DNA melting model. But, we change the concept of
$y_i$ to the $\pi$-orbital of the nucleotide. So, the short-ranged part
in Eq. \ref{Morse01} is not longer relevant. In order to simplify the
model, one may suggest to eliminate the harmonic oscillator too in Eq.
\ref{Morse01}. However, the harmonic oscillator is important for
describing the stacking interaction in the Langevin equation,
Eq.\ref{selfconsist01}. On the other hand, the bending and the torsion
of the double chain have influence over the DNA chain
\cite{Dauxois-1993}, but this behavior cannot be explained by Eq.
\ref{Morse01}, because we have just short-range exponential interactions
and a harmonic oscillator between two neighbor base pairs. The missing
long-ranged element in Eq. \ref{Morse01} is object of further research.
Finally, we remind that we do not observe the presence of the electron
in purine sequences with one replaced pyrimidine or vice versa:
$(T)_{63}$ with one $C$ in $i=32$ or $(C)_{63}$ with one $T$ in $i=32$.
The presence or absence of electrons depend of neighbor base pairs.

\begin{figure}[th]
    \epsfxsize =\textwidth
    \begin{center}
        \leavevmode
        \epsfbox{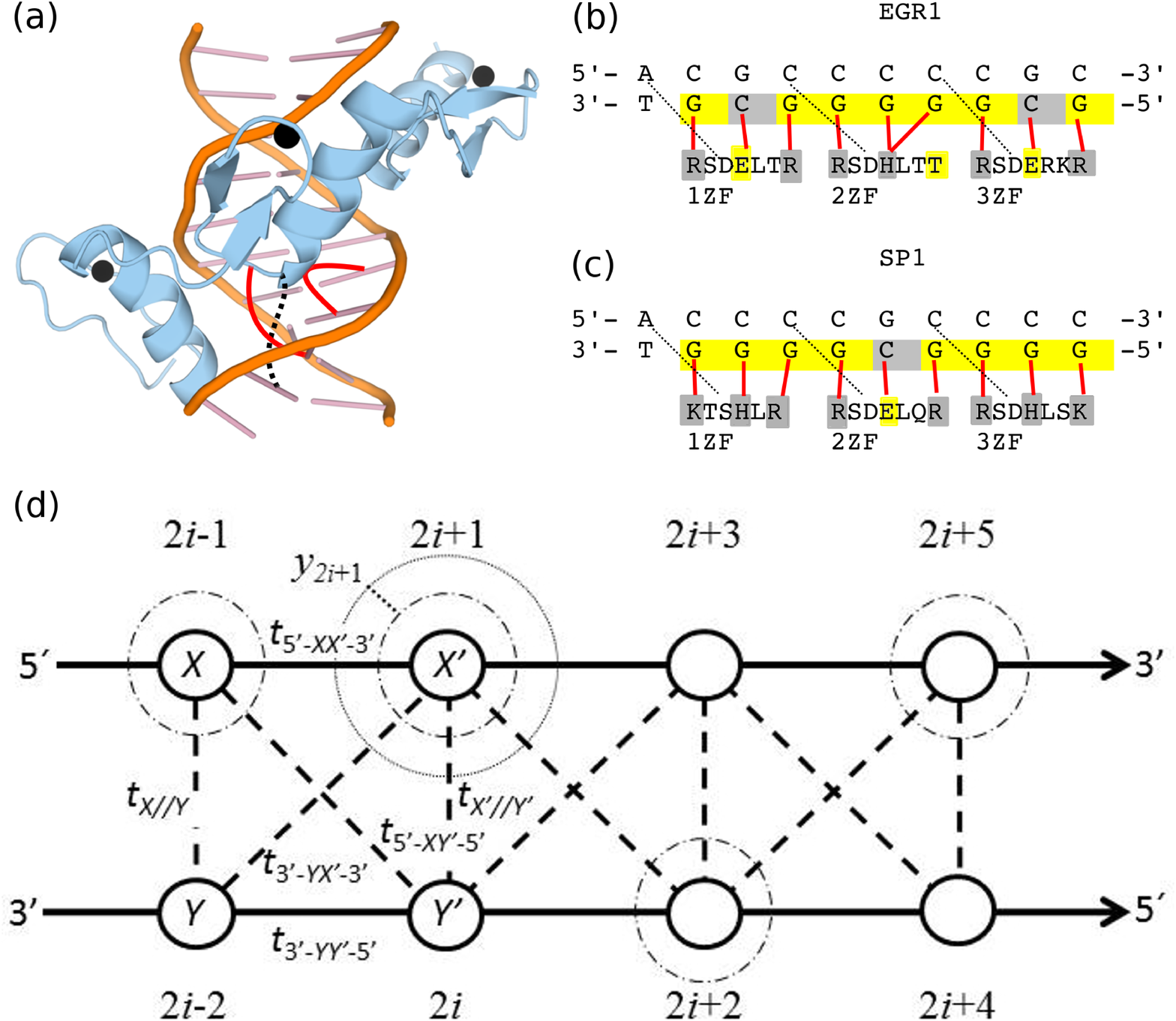}
    \end{center}
    \caption{
(a) Spatial structure of the three EGR1 zinc fingers (blue)
embracing the DNA major grove (orange). The zinc ions are in black and
the DNA-protein bindings of the second zinc finger are in red. (b) and (c) are the DNA binding
sites and amino acid sequence in the three EGR1 \cite{Wolfe-1999} and
SP1 zinc fingers \cite{Klug-2010} (1ZF to 3ZF). Solid red lines indicate
the binding between one particular nucleotide and its correspondent
amino acid. The dotted lines in (a), (b) and (c) are hydrogen bonds that
stabilize the first G-R or G-K  bonds in each zinc finger.
The nucleotides in
yellow are those with 100 \% probability of electronic presence,
when the valence band is filled, $n_e=n$. The negative charged amino
acids with weak (threonine, T) or strong acid property (glutamic acid,
E) are indicated in yellow too. The positive charged basic argine (R),
histidine (H) and lysine (K) as well as protein-binding cytosines are
indicated in gray. (d) The diagram for the DNA extended ladder model.
The light dotted line is the electronic equilibrium radius for the Morse
potential. The dark dotted lines is the field displacement $y_i$. The
dashed-dotted are the purines (adenine and guanine) electronic clouds
with $n_i=1.0$ and $n_e=n$. The dashed lines are the interstrand
electronic hopping. The solid lines are the sugar phosphate backbones.}
     \label{modelo01}
\end{figure}

The presence of electrons in purines have a profound impact in the ZF
binding. We show the consensus nucleotide sequence and the core zinc
finger binding amino acids in Fig.\ref{modelo01}(b) and (c) for EGR1 and
SP1, respectively.

The EGR1 amino acid sequence is available in the Universal Protein
Resource databank (UniProt) with the accession code P18146
\cite{UniProt-2014}. The three zinc fingers of the human EGR1 are
positioned between position 338 to 362, 368 to 390 and 396 to 418 of the
543 long amino acid sequence \cite{UniProt-2014}. The dotted lines in
Fig. \ref{modelo01}(b) are the hydrogen bonds between aspartic acid (D)
and adenine or cytosine, which stabilize the first guanine-argine(R)
hydrogen bond of ZF. The positive charged basic argine (R), histidine
(H) and lysine (K) responsible for the DNA-protein are highlighted in
gray, while the negative charged weak acid threonine (T) and strong acid
glutamine (E) are in yellow. Each red line in Fig. \ref{modelo01}(b) is
the binding of one particular nucleotide with its respective opposite
charged amino acid of the core zinc finger segment of the
EGR1\cite{Wolfe-1999}. 

The 785 amino acid long SP1 transcriptor factor, UniProt accession code
P08047, has three tandem ZFs between 626 to 650, 656 to 680 and 686 to
708 
\cite{UniProt-2014,Kaczynski-2003}. The dotted lines are the hydrogen
bond that stabilize the first ZF guanine-argine(R) or guanine-lysine(K)
bonds. The binding between core ZF amino acids and the correspondent
nucleotides are indicated by red lines in Fig.\ref{modelo01}(c)
\cite{Klug-2010}. Again, each nucleotide is connected with opposite
charged amino acid .

Concerning the electrical charges of the SP1 zinc finger tips, the
middle amino acid that bonds with the middle nucleotide of the triplet,
there is one motif associated with the position of {\color{black} BOMOs},
described in the previous section. The pattern of positive and negative
charges along the nucleotide sequence coincide with the position of the
three ZF tips. Since {\color{black} BOMOs} 
are the most stable
electrons, they guide the fingers as holder for fixing SP1 to the dDNA.
We observe the same phenomenon for the EGR1.

When we compare the strand independence analysis of the previous section
with the electronic strand dependence, one may suggest the existence of
a contradiction between the presence of {\color{black} BOMO} 
in the
complementary strand 3'-ccc-5' at EGR1 in Fig. \ref{moabegr}(h).
Actually {\color{black} BOMO} 
in this case is at the direct strand
5'-ggg-3'. We have the impression that {\color{black} BOMO} 
is in the
3'-ccc-5', since we sum the electronic cloud of direct and complementary
strand in the previous section, seeking the electronic motif of {\color{black} BOMO}.

The collective probabilities $n_i$ are not the unique strand dependent
variable in SP1 and  EGR1. The field displacement $y_i$ of the Morse
potential is also strand dependent. This means the electronic cloud
$y_i$, given by the Morse potential in Eq. \ref{Morse01}, usually
contract when $n_i=1$. i.e. in the presence of purines. The contraction
of the electronic cloud is more intense in adenine ($y_i=-0.125\pm0.001
\AA$) than guanine ($y_i=-0.114\pm0.001 \AA$). The simultaneous
measurement of the size of the electronic cloud of the direct and
complementary strands $y_i$ mirror the nucleotide order and may lead to
a new sequencing method. 

The consensus sequences, the light and dark green lines in Figs.
\ref{moabegr}(g,h),  reflect in $y_i$ and $n_i$, Figs.
\ref{moabegr}(i-l). We usually observe the  absence of the electronic
cloud in the middle cytosine of the direct strand of the SP1 and EGR1
binding sites, black lines with circle in Figs. \ref{moabegr}(i,k), as
well as the opposite behavior in the complementary strand, red lines
with plus in Figs. \ref{moabegr}(i,k). But, we should be cautious about,
because this is not true for the purine sequences with one replaced
pyrimidine or {\it vice versa} in the same way of $n_i$, as mentioned
before.

%%%%%%%%%%%%%%%%%%%%%%%%%%%%%%%%%%%%%%%%%%%%%%%%%%
%
% CONCLUSION
%
%%%%%%%%%%%%%%%%%%%%%%%%%%%%%%%%%%%%%%%%%%%%%%%%%%
\section*{Conclusion}

In the extended ladder model proposed in this article, the Morse
potential 
{\color{black} is} the key components for the
electronic behavior in the double helix DNA chain. {\color{black} But, the} stacking
interaction between adjacent base pairs in the Zhu et al.
\cite{Zhu-2007} has limited influence on the results, since this
interaction is short-ranged.

{\color{black} BOMO,} 
HOMO and LUMO show an electronic
motif behind the SP1 and EGR1 binding sites, compatible with the
consensus multiple alignments. In the case of SP1,
there is one {\color{black} BOMO} 
in the first and another in the third
zinc finger binding site, and the HOMO and LUMO positions are before the
consensus sequence. The first {\color{black}  BOMO is distributed for EGR1}
over the
second zinc finger binding position and the second {\color{black} BOMO} 
is after the
consensus sequence. The HOMO and LUMO are over the second {\color{black} BOMO.} 
{\color{black} BOMOs} 
are degenerated with
$7.98\pm0.05$eV and $7.99\pm0.03$eV for SP1 and EGR1, respectively. The
HOMO eigenvalues are $8.52\pm0.02$eV (SP1) and $8.52\pm0.01$eV (EGR1).
The LUMO energy levels are $9.3\pm0.1$eV (SP1) and $9.4\pm0.1$eV (EGR1).

When the valence band is filled, we observe a 100\%
probability in {\color{black} electronic presence} 
in purines (adenine and guanine) and
its absence in pyrimidines (thynine and cytosine). Furthermore, the
sequence of electrons and holes coincide with the basicity and acidity
of the DNA-protein binding animo acids in the zinc fingers. In
particular, the sequence of positive and negative charges of the tips of
SP1 and EGR1 coincide with {\color{black} BOMO cloud distribution.} 
Finally, 
the collective electronic behavior for the filled valence band DNA
chain will result in a sequence of electronic clouds around purine
$\pi$-orbitals, dashed-dotted lines in Fig. \ref{modelo01}(d).

%%%%%%%%%%%%%%%%%%%%%%%%%%%%%%%%%%%%%%%%%%%%%%%%%%
%
% ACKNOWLEDGMENTS
%
%%%%%%%%%%%%%%%%%%%%%%%%%%%%%%%%%%%%%%%%%%%%%%%%%%
\section*{Acknowledgments}

The authors wish to thank 
Lei Liu for the discussions about zinc fingers and for kindly providing us Fig. \ref{modelo01}(a). This work is supported by 
Conselho Nacional de  Desenvolvimento Cient\'{i}fico e Tecnol\'ogico (CNPq), process number 248589/2013, Brazil.

%%%%%%%%%%%%%%%%%%%%%%%%%%%%%%%%%%%%%%%%%%%%%%%%%%
%
% REFERENCES
%
%%%%%%%%%%%%%%%%%%%%%%%%%%%%%%%%%%%%%%%%%%%%%%%%%%

 \vfill\eject
 
 \appendix

\section{The zinc-fingers in SP1 and EGR1}

The Cys$_2$His$_2$ zinc finger unit is composed by one zinc ion between
two $\beta$-sheet and one $\alpha$-helix \cite{Wolfe-1999}, Fig.
\ref{modelo01}(a). There are two cysteine at one end of the
$\beta$-sheet and two histidines in the C-terminal portion of
$\alpha$-helix \cite{Wolfe-1999}. Each ZP bonds with three nucleotides,
Fig. \ref{modelo01} (b,c). Transcription factors use typically between 2
to 4 ZFs for identifying specific sites along the DNA 
\cite{Wolfe-1999}. In particular, the  EGR1 interact with 1-3ZF,
fitting
itself in the major groove of the dDNA, Fig. \ref{modelo01}(a). The
interaction in situ of each Cys$_2$His$_2$ zinc finger of the mouse 
EGR1 with the double DNA chain (dDNA) is detailed in
\cite{Pavletich-1991,Erickson-1996}. The EGR1 gene is a nuclear protein,
related to the cell differentiation and mitogenisis and is localized in
the position 5q31.1 of the human cytogenetic map. SP1 acts in a large
number of CG rich promoters dispersed along the genome,
\cite{Kaczynski-2003}. This 
ZF transcription factor acts as
enhancer and histone deacetylase 
binder, increasing the strength of the
chromatin wrapping by histones, and modulates DNA-binding specificity.
The gene responsible for SP1 is in 12q13.1. Here we focus on
transcription factors with few zinc fingers as for those with many ZF as
transcription the mechanism of binding is not well-understood
\cite{Klug-2010,Wolfe-1999,Lei-2015}.

\section{The model parameter choices}

When we transform the uni-dimensional Zhu et al. model \cite {Zhu-2007}
in extended ladder, each site will represent one nucleotide, instead of
a base pair. We also alter the concept of the displacement field $y_i$
\cite{Zilly-2010,Sarmento-2012}. The displacement field in the DNA
melting model is the ratio of the electronic cloud from the hydrogen
bridge between two nucleotides  \cite{Zhu-2007}. But, now the
displacement field will be the variation of the electronic cloud ratio
of the nucleotide $\pi$-orbital. So, according to the density functional
theory (DFT) studies in the nucleotides \cite{Richardson-2004,Gu-2012},
$D_A$, $D_T$, $D_C$ and $D_G$ in the Morse potential $H_b$, Eq.
\ref{Morse01}, are respectively 0.25eV, 0.44eV, 0.33eV and 0.45eV. The
DFT analysis, supported by experimental anion photoelectron spectra,
also suggests that $a_A$, $a_T$, $a_C$ and $a_G$ in the Morse potential
are correspondingly around 3.0$\AA^{-1}$, 3.0$\AA^{-1}$, 3.0$\AA^{-1}$
and 2.5$\AA^{-1}$ \cite{Chen-2007,Chen-2009}. The $k_v$ comes from DNA
vibrational spectra using Raman  spectroscopy \cite{Ghomi-1990} and we
fix this to 0.0125eV. The electronic hopping rates in the free electron
Hamiltonian $H_e$, Eq. \ref{He01}, are the same in the literature
\cite{Senthilkumar-2005,Mehrez-2005,Zilly-2010,Sarmento-2012}. Thus, we
use Table 1 for $t_{5'-XX'-3'}$, $t_{3'-YY'-5'}$, $t_{5'-XY'-3'}$ and
$t_{3'-YX'-5'}$. The interstrand hopping rate $t_{G//C}$ values 0.055 eV
and $t_{A//T}$ is fixed to 0.047 eV. The values for the on-site
potential $\epsilon_A$, $\epsilon_T$, $\epsilon_C$ and $\epsilon_G$ are
respectively 8.631eV, 9.464eV, 9.722eV and 8.178eV. In the electron-base
Hamiltonian $H_{eb}$, Eq. \ref{Heb01}, we may control the gap size
between HOMO and LUMO varying $\alpha_v$. Thus, we compare the gap in
our spectra with those reported in literature
\cite{Shapir-2008,Senthilkumar-2005,Mehrez-2005,Zilly-2010,Sarmento-2012, 
Wang-2004}, and we fix $\alpha_v$ to 1.0. 

\section{Selection of GenBank files and nucleotide alignment}
 
We use the DNA sequence from the human reference map, annotation release
106 (build GRCh38) \cite{Benson-2014}. The criteria for selecting the
binding sites in this work are as follows. The binding sites must have
experimental confirmation {\it in vitro}. We remark that we usually
observe many single nucleotide polymorphisms (SNPs) between the
reference map and the reported experimental samples, because the
reference map is basically a consensus sequence from 9 individuals
\cite{Lander-2000}  while the samples in experimental binding site data
belongs to one individual. Nested binding sites is a common occurrence,
but we try to avoid overpositioned binding sites in order to simplify
the search for an electronic motif. The binding site of these
transcrition factors is in the promoter, a region between 500 to 2000bp
distant from the beginning of the gene. We spot similar SP1 and EGR1
binding sites, TATA box and other structures reported in individual
samples in the GenBank reference map as well as in databanks as in the
Eukaryotic Promoter Database for SP1 and EGR1
\cite{Perier-1998,Praz-2002,Dreos-2013}. Finally, we use the nucleotides
sequences in FASTA and GenBank flat file format,  since the nucleotides
are just nucleotides with the phosphate group as we mentioned in the
introduction.

We select 16 binding sites in 10 different genes, see Table 2. We
remark that the  IL2 \cite{Skerka-1995} and TNF genes \cite{Yao-1997}
have binding sites for both SP1 and EGR1. The selected binding site
regions of the human genome have 50bp of length as MOAB SP1p,d
\cite{Wong-2002}, IL2 SP1 and TNF SP1. We consider segments 70bp long
around CDC25A SP1 \cite{Iavarone-1999} and PGR SP1p,d
\cite{Schultz-2003}. The segment around the three SP1 binding sites a, b
and c of the gene  SREBF1 (also known as SREBP-1a) has 90 bp
\cite{Zhang-2005}. We extract the sequences with 50bp of length for the
EGR1 binding sites for the genes EGR1 \cite{Thiel-2002}, SYN2
\cite{Petersohn-1995} and TP53 (aka P53) \cite{Thiel-2002}, but 80bp
long for VEGFA EGR1 binding site \cite{Mechtcheriakova-1999}. By the
way, EGR1 is a curious zinc finger protein, because this gene has also a
EGR1 binding site \cite{Thiel-2002}.
The EGR1 protein can regulate its own expression. The sequences of IL2
and TNF EGR1 binding sites are the same 50bp long as IL2 and TNF SP1. To
identify the binding regions we use BLAST \cite{McWilliam-2013}, SeqinR
3.1-3 \cite{Charif-2007}, Biostrings 2.34.1 \cite{Aboyoun-2015} and
ShortRead 1.24.0 \cite{Andres-2009}, since the genome has four different
readings: the normal direction ($5'-XX'-3'$ in Fig. \ref{modelo01}) ,
reverse ($3'-X'X-5'$), in the complementary strand ($3'-YY'-5'$) and in
the reverse-complementary direction ($5'-Y'Y-3'$). We indicate
respectively these reading direction with 'r' and 'c' in parenthesis for
reverse and complementary in Table 2.
 
Once we have selected the region of interest, we made multiple sequence
alignment using Clustal Omega, a software that organize the sequences
using hidden Markovian model for alignments and a multidimensional
embedding space for data clustering \cite{McWilliam-2013}. 

We adopt the same
notation for the nucleotide positioning in the promoter
\cite{Petersohn-1995,Iavarone-1999}. So, there is no position zero. The
counting always starts with 1 in the first nucleotide of the consensus
sequence. The first nucleotide before the nucleotide of the consensus
sequence is always -1. This way there is no position zero for
nucleotides. We adopt the position 1 as the first nucleotide of the
literature consensus sequence for SP1 and EGR1. In Table 2, the
consensus sequences are in light and dark green and they are defined by
a simple majority, i.e. when the number of alignment nucleotides is
bigger than or equal to 6 and 4 for SP1 and EGR1, respectively.
 
\newpage
%%%%%%%%%%%%%%%%%%%%%%%%%%%%%%%%%%%%%%%%%%%%%%%%%%
%
% FIGURES
%
%%%%%%%%%%%%%%%%%%%%%%%%%%%%%%%%%%%%%%%%%%%%%%%%%%
\section*{Figure Legends}

\begin{table}[ht]
   \begin{center}
      \begin{tabular}{c|cccc|cccc|cccc} \hline \hline 
&\multicolumn{4}{c|}{$t_{5'-XY-3'}=t_{3'-YX-5'}$}&\multicolumn{4}{c|}{$t_{5'-XY-5'}$}& \multicolumn{4}{c}{$t_{3'-XY-y3'}$}\\  \hline
&\multicolumn{4}{c|}{$Y$}&\multicolumn{4}{c|}{$Y$}& \multicolumn{4}{c}{$Y$}\\  
$X$  & G        & A         & C        & T         & G        & A         & C        & T        &    G        & A         & C        & T\\  \hline
G & 0.053& -0.077 &-0.114& 0.141 &  0.012& -0.013 &  0.002   & -0.009 & -0.032&-0.011 &0.022&-0.014\\
A &-0.010& -0.004 & 0.042 &-0.063 & -0.013&  0.031 & -0.001   & 0.007&   -0.011& 0.049 &0.017 &0.007\\
C & 0.009& -0.002 & 0.022 &-0.055 &  0.002& -0.001 &  0.001   & 0.0003 &  0.022 &0.017 &0.010&-0.004\\
T & 0.018& -0.031 &-0.028 & 0.180 & -0.009&  0.007 &  0.0003 & 0.001 &   -0.014&-0.007&-0.004 &0.006\\ \hline \hline
      \end{tabular} 
   \end{center}
   \caption{Hopping rates in eV for the extended ladder model reported in \cite{Senthilkumar-2005,Zilly-2010,Sarmento-2012}.}
   \label{txy}
\end{table}

\vfill\eject

\begin{figure}[th]
     \epsfxsize =\textwidth
     \begin{center}
          \leavevmode
          \epsfbox{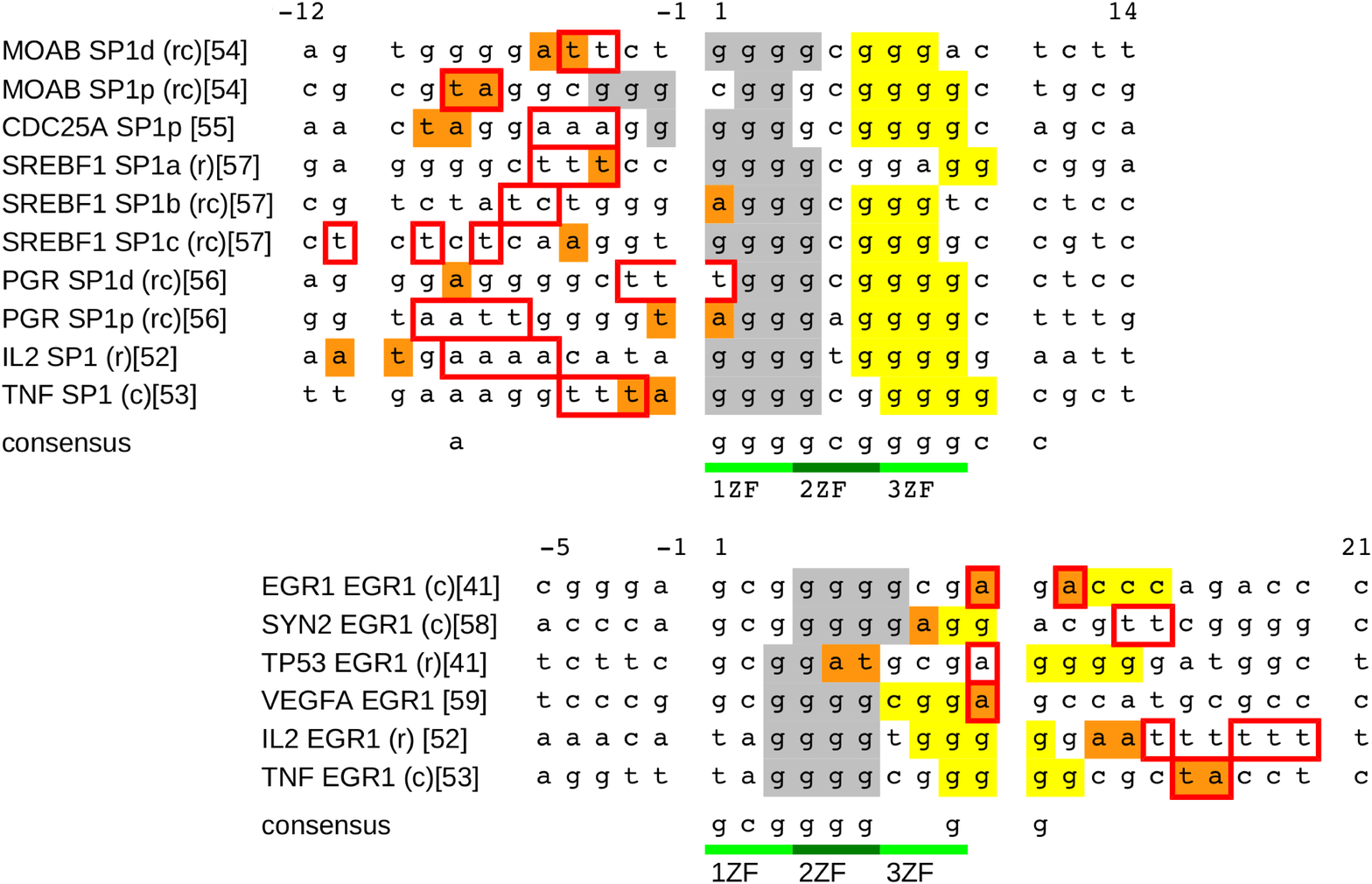}
     \end{center}
     \label{tabela11}
\end{figure}

\hspace{-0.55cm}{\bf Table 2.} 
Nucleotide alignment for SP1 and EGR1. The reading direction in reverse
and complementary strands are respectively indicated with r and c in the
parentheses. Nucleotides with at least 10\% of probability of {\color{black} electronic presence in the
bottom of the occupied molecular orbitals (BOMOs),}
$|\phi_i^0|^2\geq0.1$, are in gray
and yellow. The nucleotides with $|\phi_i^k|^2\geq0.1$ for HOMO and LUMO
are respectively indicated by orange and red bordered boxes. The consensus
sequence is the simple majority (alignment nucleotides is $\geq$ 6 and 4
respectively for SP1 and EGR1). The three zinc finger binding sites for
SP1 and EGR1  (1-3ZF) are indicated in light and dark green
\cite{Klug-2010,Thiel-2002,Skerka-1995,Yao-1997, Wong-2002,
Iavarone-1999,
Schultz-2003,Zhang-2005,Petersohn-1995,Mechtcheriakova-1999,Wolfe-1999}.

\end{document}